\documentclass[%
 reprint,
 amsmath,amssymb,
 aps,
]{revtex4-2}
\usepackage{graphicx}
\usepackage{dcolumn}
\usepackage{bm}

\usepackage[utf8]{inputenc}
\usepackage[T1]{fontenc}
\usepackage{mathptmx}
\usepackage{etoolbox}
\usepackage{gensymb}
\usepackage[colorlinks=true, urlcolor=blue, linkcolor=red, citecolor=blue]{hyperref}

\begin{document}

\title{Removal of Lunar Dust Simulant from Cold Dielectric Surfaces with Electron Beam}
\author{Hsin-yi Hao}
\affiliation{Jet Propulsion Laboratory, California Institute of Technology, Pasadena, California 91109}
\author{Wousik Kim}
\affiliation{Jet Propulsion Laboratory, California Institute of Technology, Pasadena, California 91109}
\author{David S. Shelton}
\affiliation{Jet Propulsion Laboratory, California Institute of Technology, Pasadena, California 91109}
\author{Inseob Hahn}
\affiliation{Jet Propulsion Laboratory, California Institute of Technology, Pasadena, California 91109}
\author{Benjamin Farr}
\affiliation{Laboratory for Atmospheric and Space Physics, University of Colorado, Boulder, Colorado 80303}
\affiliation{NASA/SSERVI’s Institute for Modeling Plasma, Atmospheres and Cosmic Dust, Boulder, Colorado 80303}
\author{Xu Wang}
\affiliation{Laboratory for Atmospheric and Space Physics, University of Colorado, Boulder, Colorado 80303}
\affiliation{NASA/SSERVI’s Institute for Modeling Plasma, Atmospheres and Cosmic Dust, Boulder, Colorado 80303}

\begin{abstract}
It has been demonstrated that lunar dust simulant can be efficiently lofted and removed from various room temperature surfaces in vacuum when exposed to a low-energy electron beam. This provides a potential solution to the well-known dust risks associated with future lunar exploration. Considering its application in extremely cold regions on the Moon, we experimentally demonstrated dust lofting from surfaces at temperatures as low as  -123$\degree C$ using an electron beam. Compared to room temperature applications, we found that the dust lofting from a glass surface slows down significantly at lower temperatures. Possible reasons are discussed. We also found that the dust lofting process can be accelerated when the electron beam energy is swept within an optimal range and rate. 

\end{abstract}

\maketitle

\section{\label{sec:one}Introduction}
Moon dust could hamper the extraordinary visions of NASA’s Artemis program. As learned from the Apollo missions, lunar dust sticks to all exploration hardware, causing damage to its surfaces. More importantly, its sharp and spiky nature leads to health risks to astronauts when inhaled. A variety of dust mitigation technologies have been developed over the past decades and could potentially improve the challenging situations \cite{Mohajer}.  Recently, we have developed a new technology using a low-energy electron beam (e-beam) to remove dust from various surfaces \cite{Farr2021,Farr2020,Farr2022}. This technology aims to non-invasively clean fine dust from sensitive surfaces with arbitrary shapes like solar panels, visors, and spacesuits, without adding special elements to those surfaces. 
  Following our previous room temperature studies of this e-beam cleaning method, we extend our experiments to study the efficacy of the technique at low temperature in this work. The temperature dependent dust lofting efficiency is of interest because the thermal environment of the Moon depends on latitude and varies greatly between day and night. This is especially relevant because the first landing site chosen for the Artemis mission, the south polar region of the Moon, has permanently shadowed craters at extremely low temperatures. This paper shows new experimental results regarding the temperature dependence of the e-beam dust lofting efficiency on glass and Ortho-fabric spacesuit sample surfaces. In addition, moisture effects of prepared lunar simulants in different bake-out conditions are also presented. 

\section{\label{sec:one}Experimental Setup}
The schematic of the experimental setup is illustrated in Figure 1a. The experiment was conducted in a bell-jar vacuum chamber about 90 cm in diameter and 90 cm tall. The dust sample assembly, which is described in Sec. 3, was mounted on a rotatable aluminum sample plate with screws and copper tape for better thermal contact. The sample plate was mounted on a support frame using a thin wall stainless steel tube as a shaft for thermal isolation. A motor was coupled to the shaft to rotate the sample plate. For low surface temperature measurements, the back of the sample plate was thermally linked to the first stage of a cryocooler cold head using braided copper strands for flexibility. The temperature of the sample could reach as low as -$123$\degree C with this arrangement. For high surface temperature measurements, a heater was used on the back of the sample plate to raise the temperature to around $80\degree C$. Dust lofting experiments were performed when the pressure of the chamber was below $5\times10^{-6} torr$. During the low surface temperature dust lofting experiments, the cryocooler acted as a cryo-pump and the chamber pressure was as low as $5\times10^{-8} torr$. 

\begin{figure}[h]
\centering
\includegraphics[width=1.0\linewidth]{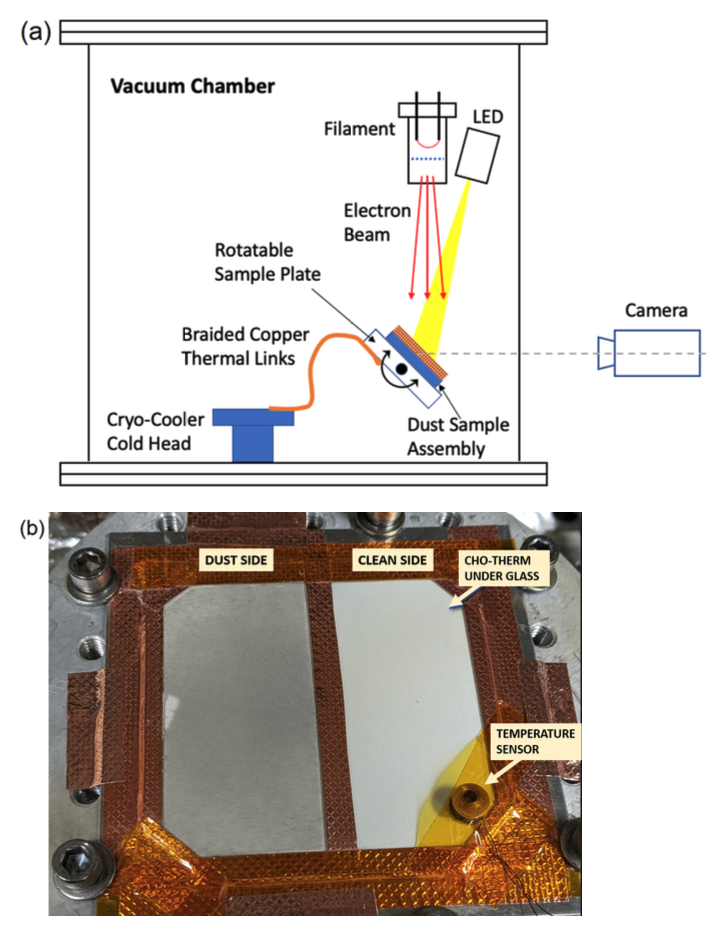}
\caption{(a) Schematic of the experimental setup inside a vacuum chamber. An electron beam produced by a negatively biased hot thoriated tungsten filament bombards the sample mounted on a rotatable sample plate, which is thermally linked to the cold head of a cryocooler by braided copper strands.  A yellow LED illuminates the sample, so that a camera outside the chamber can monitor the progress of the dust lofting. (b) An example of a sample assembly with glass surface. It consists of a dust side and a clean side. The normalized brightness of the sample is defined as the average brightness of the dust side divided by the average brightness of the clean side.}
\label{fig:figure1}
\end{figure}

  The electron beam used for dust lofting was emitted from a negatively biased hot thoriated tungsten filament mounted $25 cm$ above the dust sample surface. A grounded grid was directly below the filament to generate an electric field to accelerate the electron beam, whose energy was determined by the bias voltage at the filament. This electric field eliminates space charge effects near the filament, creating high beam currents without the need of a plasma.
 The e-beam profile had a maximum flux density near the horizontal center of the sample. The voltage across the filament sets the heating current of the filament, which determined the electron current density, $\sim100 \mu A/cm^2$, measured by a Faraday cup.  The electron beam energy was controlled to vary between 50 and 425$eV$.
  The dust sample assembly was normally mounted at $45\degree$ angle to the horizon, illuminated by a yellow LED light (Thorlabs M565L3 - 565 $nm$) from above, for a camera (Ximea xiB PCIe x4 Gen 2) to record images during the dust lofting process. 

\section{\label{sec:two}Sample Preparation}

For each test, a sample assembly was built. Two types of surface materials were used in our study. The first was a $760 \mu m$ thick uncoated ceria-doped microsheet from Qioptiq - Excelitas Technologies. It is typically used as cover glass for solar cells in space applications. An example of the glass sample assemblies is shown in Figure 1b. To provide good thermal contact and a light color background for observation, a piece of Cho-Therm was sandwiched between a $7.5 cm\times7.5 cm$ glass sample and the aluminum support plate. They were then held together with embossed 3M 1245 copper foil tape to help with heat conduction. The LED light did not illuminate the sample uniformly and its brightness also gradually decreased during the test. To help normalize the brightness of the sample, the glass was divided into a clean side on the right and a dust side on the left by a narrow strip of copper tape. Only the dust side was initially covered with dust and the clean side was used as a reference. For each recorded image, the normalized brightness of the sample, which is conveniently used to characterize the cleanliness of the sample, is defined as the average brightness of the pixels on the dust side divided by that on the clean side \cite{Farr2021,Farr2020,Farr2022}. After each test, the glass was thoroughly scrubbed to remove any hard-to-clean residual dust before we loaded the fresh dust simulant for the next trial. 
  The second surface material we tested was an Ortho-fabric spacesuit sample developed for the Space Shuttle program. The choice of the textile and the fabric construction were specifically made for EVAs in Low Earth Orbit, therefore, not suitable for lunar surface exploration. However, with no coating to protect against dust and debris, it provides a more challenging case for our cleaning technique. The fabric was also mounted to an aluminum support plate with copper tape on its periphery and double-sided tape on its back for good thermal contact. As with the glass sample, the Ortho-fabric sample was also divided into a dust side and a clean side for the brightness normalization purpose.
  The dust sample used in this study was LHS-1 Lunar Highlands Simulant from Exolith Lab. Following NASA-STD-1008 guideline for Planetary External environment, we tested the dust sample with three different bake-out conditions (Type U: unbaked, Type A: $110\degree C$ for 24 hours, and Type B: $200\degree C$ for 24 hours) to examine whether the bake-out affects the efficiency of dust lofting. Results from stationary cleaning show that it is significantly harder to loft baked dust, especially Type B, from Ortho-fabric surfaces (Fig. 2a). The effects also exist, to a lesser extent, for glass surfaces (Fig. 2b). Following the lunar simulant moisture level best practices specified in the guideline, we used Type B dust for the rest of the study. To retain the dryness of the baked dust, immediately after the bake-out, the dust samples were divided in small Amerstat bags within a nitrogen flooded glove box and stored in a sealed jar with desiccant pouches and a moisture indicator.
  Before loading the dust on the sample assembly, we minimized possible static charges on the sample assembly using an air ionizer. The dust loading was performed in the glove box with relative humidity between $3\sim10\%$ during the operation. A standard sieve with $45 \mu m$ openings was used to load the dust on the sample surface. The thickness of the dust on the sample was not controlled, but the surface was covered as evenly and consistently as possible by eye. The sample was then transferred into the vacuum chamber from the glove box within five minutes to minimize moisture reabsorption. The relative humidity of the room was not controlled and it varied between $25\%$ to $55\%$ from day to day. The sample assembly was mounted on the rotatable sample plate with bolts and copper foil tape for good thermal contact. To monitor the temperature of the sample, a Cernox thermometer was attached to a corner on the clean side of the surface without touching the copper foil tape. The chamber was pumped to a pressure below $5 \times10^{-6} torr$ before the dust cleaning. The pump-down took at least 15 hours, allowing some of the re-absorbed moisture the time to escape from the simulant.

\begin{figure}[h]
\centering
\includegraphics[width=1.0\linewidth]{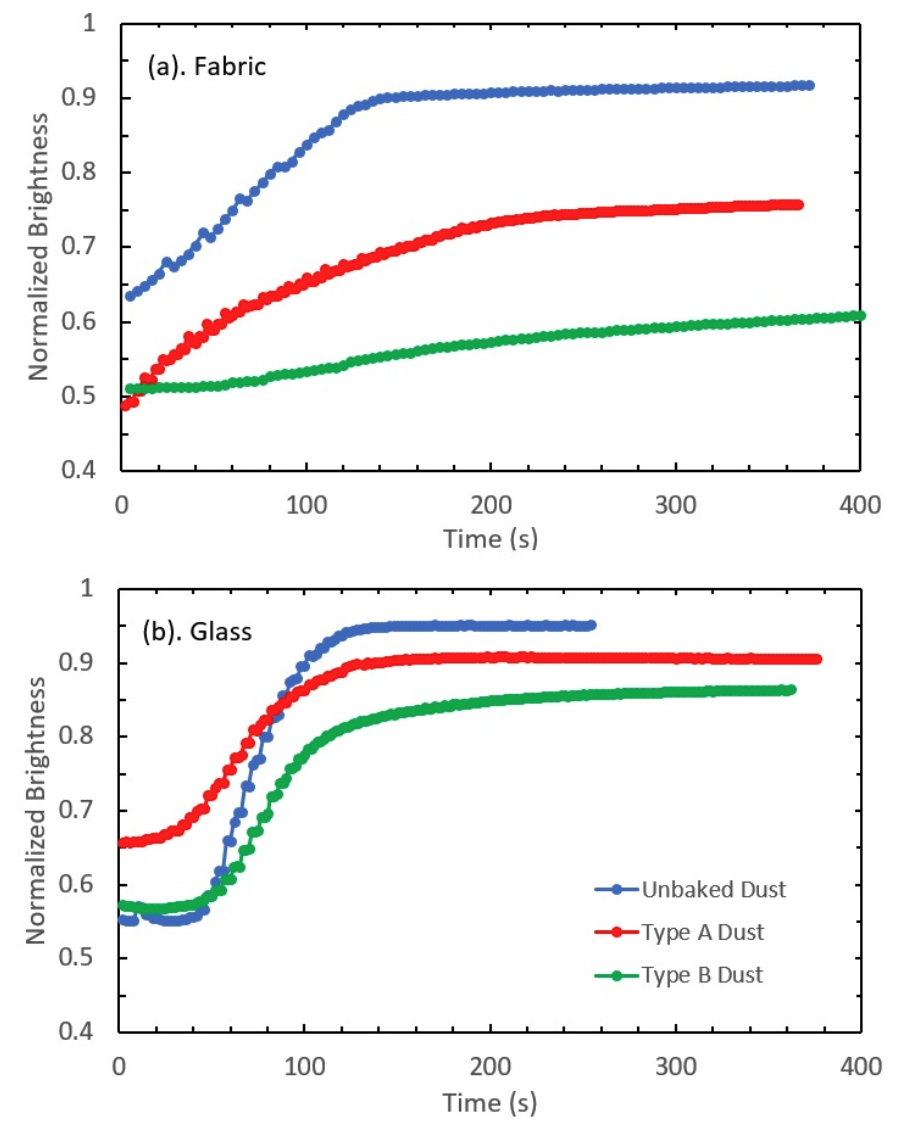}
\caption{Relative cleanliness, characterized by the normalized brightness, as a function of time for two types of surfaces (Ortho-fabric and Glass) and three different levels of dust dryness (Type U: unbaked, Type A: $110\degree C$ for 24 hours, and Type B: $200\degree C$ for 24 hours) during stationary cleaning process. The cleaning efficiency is significantly hampered by dryness of the dust sample on the Ortho-fabric surfaces, but less so for the glass surfaces. The final cleanliness is also lower for the dryer samples}
\label{fig:figure2}
\end{figure}

\section{\label{sec:three}Experimental procedure}

In a previous study \cite{Farr2021}, it was shown that varying the incident angle of the e-beam improves the final surface cleanliness of samples. In this study, we have implemented this mechanism and derived a cleaning procedure that achieved the maximum cleanliness over the shortest time. It includes two steps. First, the sample surface is held stationary at $45\degree$ relative to the electron beam and has its image taken at a frame rate of $0.25 fps$ until the cleaning rate begins to slow down. Then we start rotating the sample plate to continuously vary the e-beam incident angle between  -$36\degree$ and $45\degree$ (the range is limited by the attachment of copper thermal links), pausing periodically at $45\degree$ to take a sample image throughout the rest of the cleaning process. It was tested that no dust fell off during the rotation due to gravity alone.
  The optimal e-beam energy for dust cleaning depends on dust properties, surface material, and surface temperature. We found that sweeping the bias voltage at the filament within a fixed voltage range not only allows us to cover the optimal voltage for different properties of the dust sample and substrate surface, but it also allows us to make consistent comparison between testing trials. Figure 3 shows the effects of voltage sweeping during the dust cleaning process for a room temperature glass surface. Although the sweeping method has a slower initial cleaning rate compared to using the optimal energy of $200 eV$ for this configuration, it has overall a steadier rate and continues to be effective after the fixed energy method stops. After the subsequent rotational cleaning, all three samples reach similar final cleanliness within almost the same amount of time. For all the temperature dependent trials, we swept the bias voltage between $125V$ and $425V$ at an absolute rate of $120V/s$.
  In our study, we characterize the relative cleanliness of the sample using the normalized brightness defined in the previous section. Note that because different types of surfaces reflect light differently in the clean state, one should not directly compare cleanliness between samples with glass and Ortho-fabric surfaces. In addition, for the Ortho-fabric, the weave causes the light to be reflected at different angles across the sample, generating saturated pixels in some part of sample images. It is only practical to allow saturation in some pixels, otherwise the resulting images will be too dark to discern the change in brightness. As a result, it is important to obtain the same overall brightness on the clean side for all fabric samples to allow for valid comparisons between different sets of data. 

\begin{figure}[h]
\centering
\includegraphics[width=1.0\linewidth]{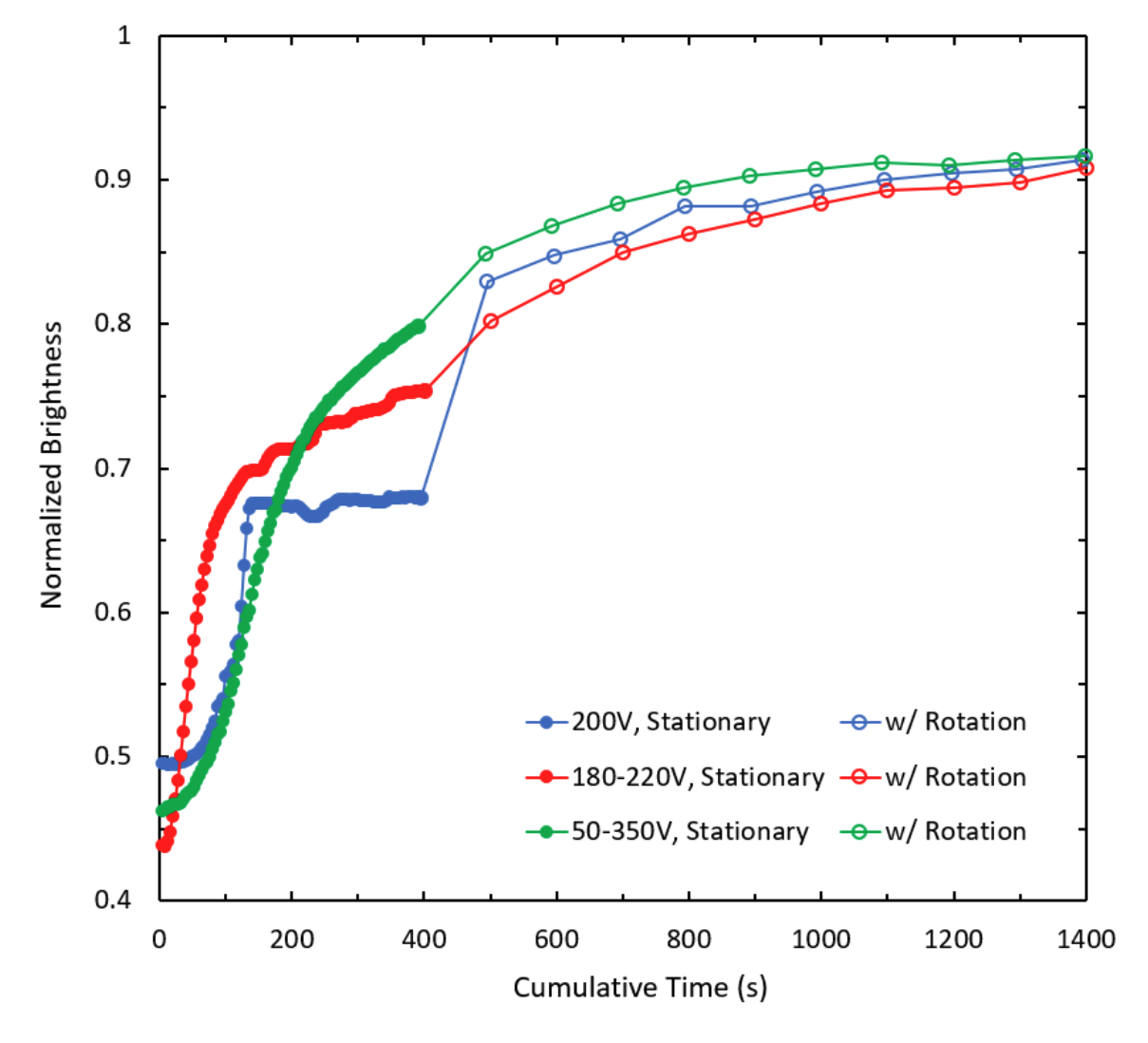}
\caption{The effects of voltage sweeping during the dust cleaning process for room temperature glass surface. During the stationary cleaning, the process with fixed bias voltage at $200 V$ stopped after around two minutes time. Meanwhile, the process with widest sweeping voltage range, from $50 V$ to $350V$, had a slower start but a steadier overall cleaning rate. However, the subsequent rotational cleaning brought all three samples to almost the same final cleanliness within about the same amount of time.}
\label{fig:figure3}
\end{figure}

\section{\label{sec:four} Results}
 Figures 4 and 5 show temperature related changes in cleaning efficiency for Type B dust on the glass and Ortho-fabric surfaces, respectively. Each trace represents a separate trial. The normalized brightness, which characterizes the relative cleanliness, is plotted as a function of cumulative cleaning time and composed of an initial stationary cleaning (closed circles $\bullet$) followed by cleaning with rotation of the sample plate to vary the incident angle of the e-beam (open circles 
 $\circ$). For the glass surfaces, compared to the room temperature result, the overall cleaning efficiency reduces significantly at a temperature as low as  $106\degree C$ and, on the other hand, slightly increases at elevated temperatures of $63\degree C$ and $80\degree C$. 
\begin{figure}[h]
\centering
\includegraphics[width=1.0\linewidth]{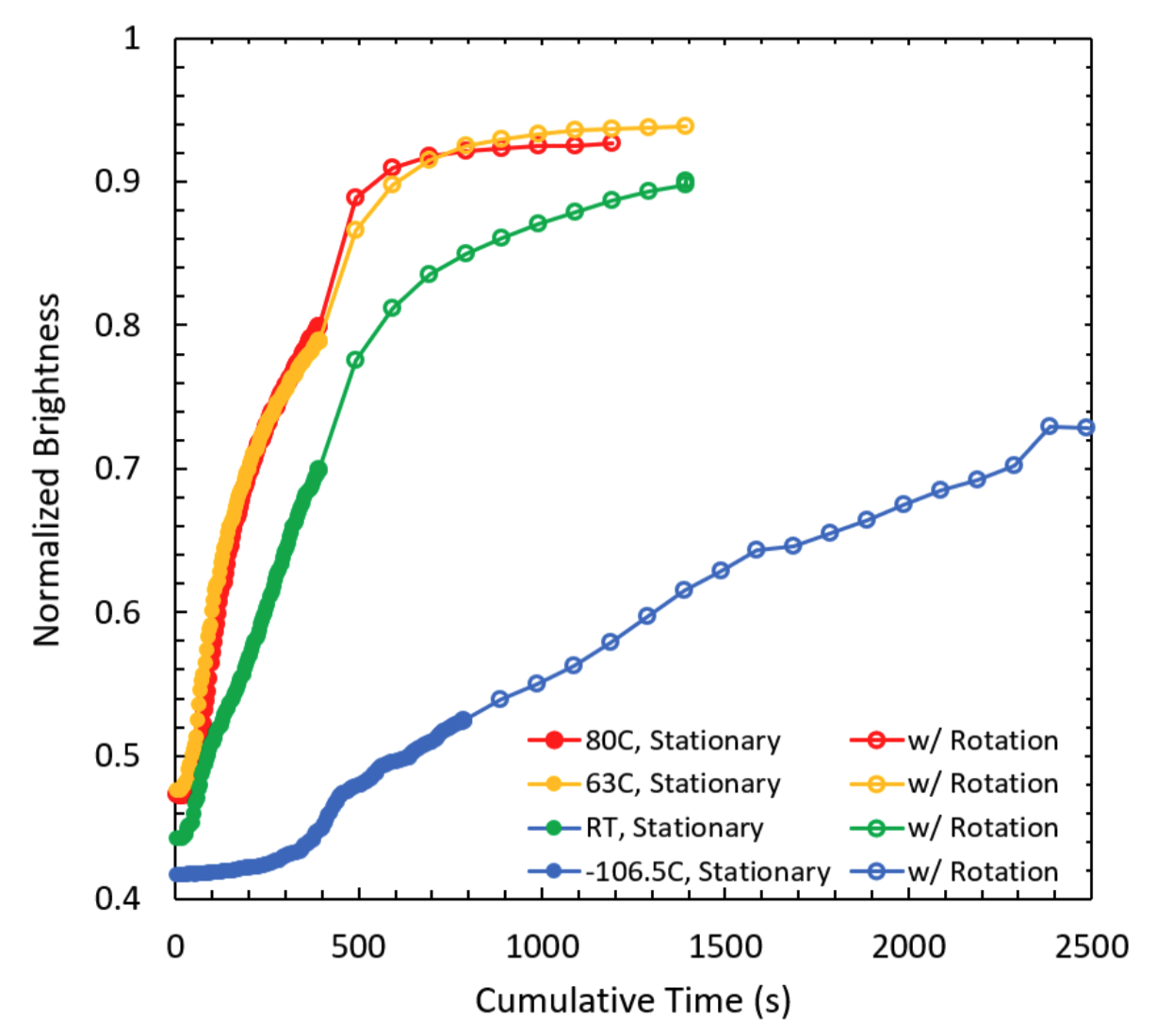}
\caption{Cleanliness as a function of time for Type B dust on glass surface at different temperature. The cleaning efficiency is significantly reduced at low temperatures and slightly improved at elevated temperatures.}
\label{fig:figure4}
\end{figure}

\begin{figure}[h]
\centering
\includegraphics[width=1.0\linewidth]{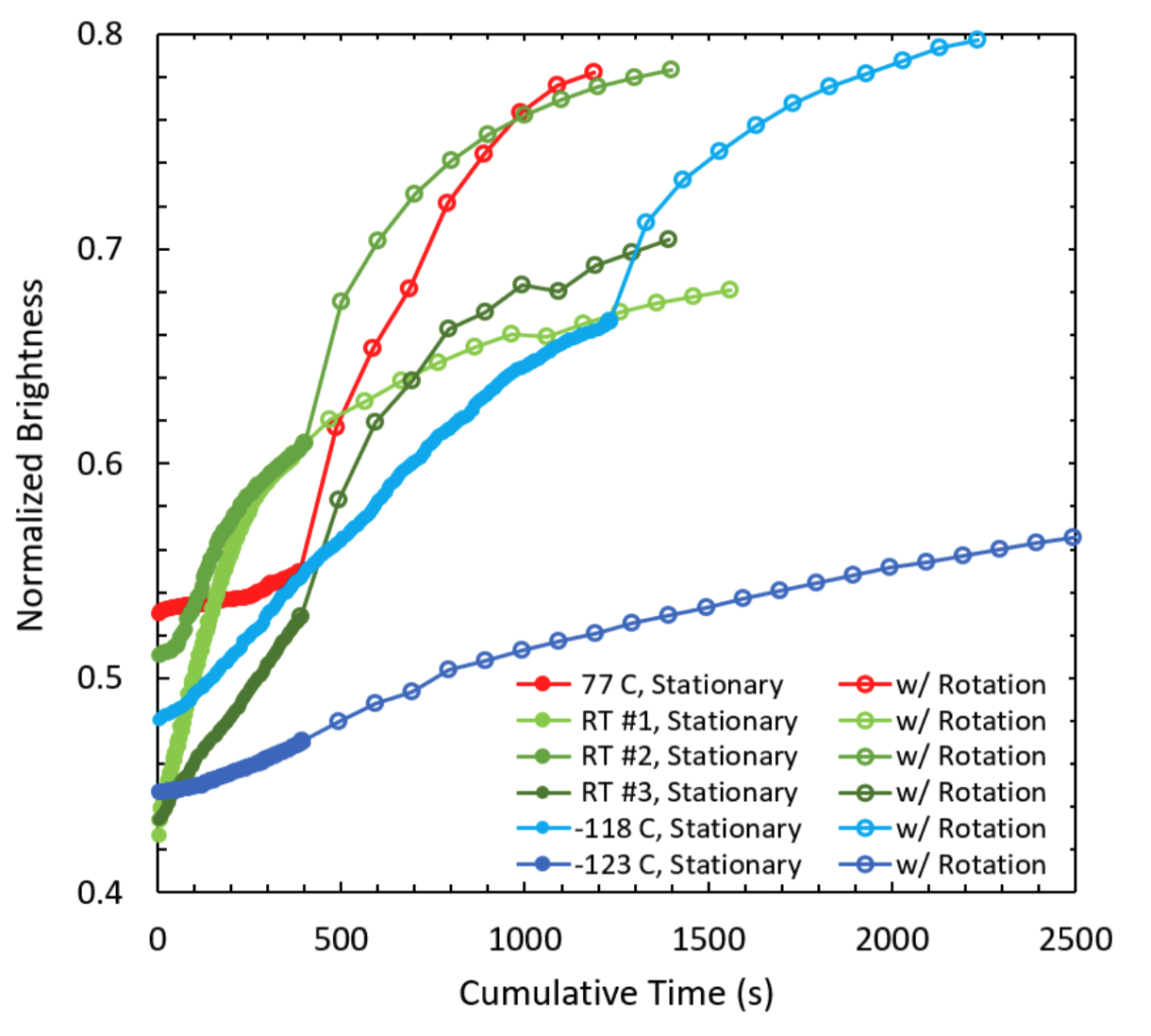}
\caption{Cleanliness as a function of time for Type B dust on Ortho-fabric surface at different temperature. The cleaning efficiency is somewhat reduced at low temperatures. Rotation of the sample increases the cleaning efficiency in general, but its effects depend on the initial loading conditions which are difficult to control in our experiments. }
\label{fig:figure5}
\end{figure}

  For the Ortho-fabric surfaces, the effects of surface temperature on the cleaning efficiency are not as clear as for the glass samples. Except for one cold sample at  -$123\degree C$ that was very hard to clean, we did not observe a reliable temperature dependence for the cleaning efficiency.  One significant observation is the random nature of how the cleaning process unfolded; even at the same temperature, e.g., the three green traces for room temperature trials in Figure 5, we have different cleaning results. In some cases, dust was lofted fairly evenly across the surface and the cleaning efficiency improved after rotational cleaning started. In other cases, an avalanche quickly built up, cleaning out upper part and leaving dust to be piled up at the lower part of the surface. After that, the cleaning rate held steady and did not see a boost when switching to rotational cleaning. One possible reason for these seemingly random behaviors is that due to the surface topography of the Ortho-fabric, the dust lofting from such surfaces is highly sensitive to the thickness and uniformity of the initial dust distribution. At more heavily loaded locations, gravity pushes more dust particles into nooks and crannies of the weave, making them hard to clean. Since our dust loading was done by hand and judged by eye, the resulting different dust distribution between samples could plausibly contribute to inconsistent cleaning results. Additionally, slight variations in the moisture level in the loaded dust between trials may contribute to the uncertainty of the cleaning results as well. As shown in Figure 2a, the dryness level of the dust shows a pronounced effect on the cleaning efficiency for Ortho-fabric. 
We should also note that the e-beam profile is non-uniform, with the highest flux in the middle part of the sample, leaving more residual dust at the upper and lower ends of the surface. It is desirable to modify the e-beam source so that a more uniform flux distribution can be obtained in future applications. 

\section{\label{sec:five}Discussion}
As described in our previous work  \cite{Farr2021,Farr2020,Farr2022}, the e-beam dust cleaning technique is based on the patched charge model \cite{Wang2016}. In this model, an e-beam could enter in openings of microcavities formed between dust particles covering a surface and back-scatter from the dust particles underneath and/or penetrate the particles, generating secondary electrons. The released low-energy secondary electrons could travel to adjacent particles within the microcavity and build up surface charges. When the resulting Coulomb forces are larger than the cohesive forces between particles and the adhesive forces between particles and surface, particles are detached from each other and lofted from the surface. 

To understand the temperature dependence of dust lofting efficiency observed in our experiment, we have looked into potential sources of temperature dependence in van der Waals (vdW) forces, which are mainly responsible for the cohesion/adhesion forces among particles and surface, as well as temperature dependence in the secondary electron yield. Our discussion of apparent temperature effect on dust lofting only applies to the simple flat glass surface. For the case of Ortho-fabric samples, the interpretation of its temperature dependent lofting behavior is complicated by the nature of the woven fabric structure where many dust particles could be mechanically trapped in pockets.

\subsection {Temperature dependent London-van der Waals force}

According to the Lifshitz theory of vdW forces between macroscopic bodies, as two media filling half spaces with plane-parallel boundaries separated from one another by a distance $d$, temperature effects are not important at small separations $(d\ll \hbar c/(k_B T))$, such that $F_{vdW}\sim\hbar \tilde\omega/d^3$, where $T$ is the absolute temperature of the bodies, $d$ is the separation, $k_B$ is the Boltzmann constant, $\hbar$ is the reduced Planck constant, $c$ is the speed of light, and $\tilde\omega$ is a frequency characterizing the adsorption spectra of the two bodies \cite{Landau}. In the opposite limit, where the separation is large, the attractive force depends on temperature, so that $F_{vdW}\sim(k_B T)/d^3$ .  Strictly speaking, these simple formulas are derived for limited cases and for an ideal geometry. However, the main temperature-dependent term in the theory is originated from statistical treatments of electromagnetic fluctuations of a macroscopic body at a finite temperature, independent of the geometry. For instance, the temperature-dependent vdW force between two spheres is also studied theoretically, and its temperature dependent leading term is essentially the same \cite{Mitchell}. Our sample/dust geometry is more complex; however, a general trend in temperature dependency can still be inspected qualitatively with the Lifshitz theory. 
In our experiments, the LHS-1 lunar dust simulant has a particle size distribution approximately from $0.1\mu m$ to $45\mu m$. While it is difficult to obtain the statistical distribution of the distance, $d$, between two adjacent particles or between the dust particle and the substrate surface, it is very likely that $d$ is in general smaller than the diameter of the largest particle in our sample. Hence, the Lifshitz’s criterion $(d\times k_B T)/\hbar c$ is less than 1 in our experimental conditions. Therefore, we anticipate the cohesive/adhesive forces in our dust/surface system being nearly temperature independent. Since our experimental observations have shown that it is more difficult to loft dust particles from glass surfaces at low temperatures, the van der Waals force cannot explain the temperature-dependent behavior of the dust lofting efficiency for our simple dust/glass systems. 

\subsection{Temperature dependent charging effects}

As described above, the essence of the patched charged model is that secondary electrons escaped from the surfaces of dust particles exposed to an e-beam and accumulated on the surfaces of neighboring particles within the microcavities \cite{Wang2016}. A basic semi-empirical theoretical framework to understand the universal nature of the secondary electron emission from a material under e-beam exposure is well established \cite{Bruining,Dekker}.  In general, high energy primary electrons generate cascades of secondary electrons inside the material, but these secondary electrons do not always escape from the surface of the material. Instead, the secondary electrons lose their energy through collisions with other electrons and interactions with lattice. These internal energy loss mechanisms are different for metals and insulators. In insulators, as in our lunar simulant particles, there are fewer conduction electrons (less collision events) than in metals, and the secondary electrons lose much less energy as they move through such materials. This is why insulators typically show a larger secondary electron yield (SEY) compared to metals, as they allow secondary electrons to travel farther to reach the surface from the creation sites inside the materials \cite{Shih}.  Secondary electrons interact with lattice vibrations, which are temperature dependent, resulting in temperature dependent SEY. Such interactions are more significant in insulators than in metals, where electron-electron interactions dominate. The temperature dependent SEY in insulators has rarely been studied experimentally and we have only found a limited number of such work in literature. In studies for a wide temperature range (from room temperature to $740\degree C$, way beyond the temperature range of our experiments), it was shown that elevation of the temperature resulted in a decrease in SEY. Those results were attributed to the interaction of secondary electrons with higher level of lattice vibrations (phonons) at higher temperature \cite{Johnson,Belhaj}.  In a relatively narrow temperature range, Balcon et al. observed an increase of SEY from silicate glass when the temperature increased from room temperature to $80\degree C$ \cite{Balcon}, a result seemingly contradicting the above findings for higher temperature ranges. And this observation is consistent with temperature dependent conduction mechanisms (e.g., diffusion, Poole-Frenkel de-trapping, and hopping conduction) of electrons inside insulating materials, which generally predict an enhanced electrical conductivity when temperature increases \cite{Zallen,Ambegaokar}. As a result, the increased conductivity makes secondary electrons to flow upward to the surface more easily, causing an increase in the SEY as described \cite{Balcon}. Our experimental results show the temperature-enhanced dust lofting efficiency between  -$110\degree C$ and $80\degree C$. Although our experiment does not measure the SEY directly, the observed temperature dependent dust lofting efficiency is consistent with reduced SEY at lower temperatures.
It should be noted that for different dust cleaning methods, the temperature dependence of the cleaning efficiency could be different due to different mechanisms employed. For example, a study by Kawamoto and Hashime, using an electrostatic traveling wave, found no performance difference between -$25\degree C$ and $120\degree C$ in vacuum \cite{Kawamoto}.

\section{\label{sec:six}Summary}

We have found that the cleaning performance of the e-beam dust removal technology is reduced for some of the surface/dust systems at cryogenic temperatures below  -$100\degree C$. Although it is anticipated that most of human engineered surfaces’ demand of dust removal on the Moon is in sunlit conditions at higher temperatures (e.g., solar panels, spacesuits, visors, and thermal radiators) or in an airlock at room temperature, there will be needs for dust cleaning of low temperature surfaces, especially for the exploration hardware in permanently shadowed areas on the Moon. Our future works will address these additional challenges to improve the efficiency of the e-beam technology at extremely low temperatures and the cleaning of last layers of dust particles on the surface. Finally, we would like to point out the importance of demonstrating the dust lofting effectiveness of the e-beam technology (or even other technologies) on the Moon in advance. This is because the Moon dust cohesion/adhesion characteristics could strongly depend on the Moon’s natural environments (e.g., thermal, charging, local plasmas, electromagnetic fields, high vacuum, and the actual physical properties of the dust), which are not practical to simulate precisely on Earth \cite{Walton}.

\section{\label{sec:six}Acknowledgment}

  This research was carried out at Jet Propulsion Laboratory, California Institute of Technology, supported by NASA/BPS and Game Changing Development Dust Mitigation project through a contract 80NM0018F0506, and by the NASA/SSERVI’s Institute for Modeling Plasma, Atmospheres and Cosmic Dust (IMPACT). We would also like to thank Evelyne Orndoff of NASA Johnson Space Center for providing the Ortho-fabric, Christopher Wohl of NASA Langley Research Center for the LHS-1 Lunar Highlands Simulant, and Joel Schwartz of Jet Propulsion Laboratory for the solar panel cover glass used in this study.

\bibliography{bib}
\end{document}